\documentclass[twocolumn]{aastex62}

\usepackage{commath} 
\usepackage{gensymb} 
\usepackage{eucal} 
\usepackage{aas_macros}
\usepackage{graphicx}
\usepackage[caption=false]{subfig} 
\usepackage{natbib}
\usepackage{hyperref}

\usepackage{color}

\renewcommand{\vec}[1]{\boldsymbol{#1}}

\newcommand{\uvec}[1]{\hat{\vec{#1}}} 
\newcommand{\mat}[1]{\boldsymbol{\mathbf{#1}}}
\newcommand{\RNum}[1]{\uppercase\expandafter{\romannumeral #1\relax}}

\DeclareMathOperator{\rank}{rank}

\DeclareMathOperator{\Tr}{Tr}

\newcommand{\MHz}{\ensuremath{\, {\rm MHz}}}

\begin{document}

\title{An Eigenvector-based Method of Radio Array Calibration and Its
Application to the Tianlai Cylinder Pathfinder}


\author{Shifan Zuo}
\affiliation{Key Laboratory for Computational Astrophysics, National Astronomical Observatories, Chinese Academy of Sciences, Beijing 100101, China}
\affiliation{School of Astronomy and Space Science, University of Chinese Academy of Sciences, Beijing 100049, China}

\author{Ue-Li Pen}
\affiliation{Canadian Institute for Theoretical Astrophysics, 60 St. George Street, Toronto, Ontario M5S 3H8, Canada}
\affiliation{Canadian Institute for Advanced Research, CIFAR Program in Gravitation and Cosmology, Toronto, Ontario M5G 1Z8, Canada}

\author{Fengquan Wu}
\affiliation{Key Laboratory for Computational Astrophysics, National Astronomical Observatories, Chinese Academy of Sciences, Beijing 100101, China}

\author{Jixia Li}
\affiliation{Key Laboratory for Computational Astrophysics, National Astronomical Observatories, Chinese Academy of Sciences, Beijing 100101, China}
\affiliation{School of Astronomy and Space Science, University of Chinese Academy of Sciences, Beijing 100049, China}

\author{Albert Stebbins}
\affiliation{Fermi National Accelerator Laboratory, Batavia, IL 60510, USA}

\author{Yougang Wang}
\affiliation{Key Laboratory for Computational Astrophysics, National Astronomical Observatories, Chinese Academy of Sciences, Beijing 100101, China}

\author{Xuelei Chen}
\correspondingauthor{Xuelei Chen}
\affiliation{Key Laboratory for Computational Astrophysics, National Astronomical Observatories, Chinese Academy of Sciences, Beijing 100101, China}
\affiliation{School of Astronomy and Space Science, University of Chinese Academy of Sciences, Beijing 100049, China}
\affiliation{Center of High Energy Physics, Peking University, Beijing 100871, China}
\email{xuelei@cosmology.bao.ac.cn}

\begin{abstract}
  We propose an eigenvector-based formalism for the calibration of radio
  interferometer arrays. In the presence of a strong dominant point source, 
  the complex gains of the array can be obtained by taking the first eigenvector 
  of the visibility matrix. We use the stable principle component analysis (SPCA) method
  to help separate outliers and noise from the calibrator signal to improve the performance of the method. 
  This method can be applied with poorly known beam model of the antenna,
  and is insensitive to outliers or imperfections in the data,
  and has low computational complexity. It thus is particularly suitable
  for the initial calibration of the array, which can serve as the
  initial point for more accurate calibrations.  We demonstrate this method by applying it to the
  cylinder pathfinder of the Tianlai experiment, which aims to measure the dark
  energy equation of state using the baryon acoustic oscillation
  (BAO) features in the large scale structure by making intensity
  mapping observation of the redshifted 21~cm emission of the
  neutral hydrogen (HI). The complex gain of the array elements 
  and the beam profile in the East-West direction (short axis of the cylinder) are successfully obtained 
  by applying this method to the transit data of bright radio sources.
\end{abstract}



\section{Introduction}\label{S:intro}
Calibration of a telescope is to determine the various parameters
which characterize the telescope model by solving equations 
linking the observational data to these parameters. In the case of a radio interferometer array, 
the model typically includes the beam and polarization response, the
band pass, and the complex gain of the receiving elements.  
In most cases, even if the beam response of the telescope is relatively stable,  
the amplitudes and phases of the receivers (complex gain) still vary significantly and
must be calibrated during observation. Many interferometer array calibration methods have been developed 
and are in wide use  (see e.g. \citealt{Thompson1986,Perley1999,Sault1996,Hamaker2000,
Smirnov2011a,Smirnov2011b}).  In recent years, with the 
need of achieving high precision for arrays with very large number of elements, and especially the 
low frequency arrays which have large field of view (FoV) where direction-dependent beam response must be 
taken into account, the calibration methods  are further developed and refined, e.g. the SAGECal algorithm \cite{Kazemi2011},
the Wirtinger derivative method \cite{Tasse2014}, the Statistically Efficient and Fast Calibration (StEFCal) \cite{Salvini2014}, the Complex optimization method \cite{Smirnov2015}, the
Facet calibration method \cite{Weeren2016}, etc. 

Calibration is usually a multi-step and iterative process. 
After a reasonably good initial model of the telescope is achieved, the model is 
refined to take into account smaller effects. While the initial calibration is a coarse one, 
it also has the challenge that the model of telescope is largely unknown, so it needs to be
blind and robust. In this paper, we present a method of calibrating the complex gains of the interferometer array
based on eigenvector decomposition\footnote{This method was previously used by K. Bandura in the 
calibration of the Pittsburgh cylinder in his Ph.D. thesis \citep{2011PhDT.......158B}.}, which is accurate and computationally efficient. 
To make it more robust in the presence of missing data or occasional outliers, we also improve the method by 
using a technique called the stable principal component analysis (SPCA)  to separate the dominant calibrator signal, 
the noise and the occasional outlier components by exploiting their different properties in the covariance matrix. 
As a concrete example, the method is applied to the calibration of the Tianlai cylinder array pathfinder.

The Tianlai\footnote{\url{http://tianlai.bao.ac.cn}} (Chinese
for ``heavenly sound'') project   \citep{2012IJMPS..12..256C,Xu:2014bya} is an experimental
effort to make intensity mapping \citep{2008PhRvL.100i1303C} observations of the redshifted 
21cm line from the neutral hydrogen,
in order to measure the baryon acoustic oscillation (BAO) signal of large scale structure, and measure the 
dark energy equation of state. The Tianlai pathfinder includes both a dish array with 16 dishes, compactly 
arranged in two concentric rings \citep{Zhang:2016whm}, and a cylinder array with three north-south oriented
cylinders \citep{Zhang:2016miz}, containing 31, 32, and 33 feed elements respectively. The construction of the 
two arrays was completed in 2015, and the first trial observation were done in September 2016. 
We have developed a data processing pipeline for the arrays, and here we present the method of
its initial calibration.

This paper is organized as follows:  In Sec. 2 we introduce the
basic principle of the complex gain determination using
eigenvector analysis method and its generalization to the stable PCA
method. In Sec.3 we apply the method to the Tianlai array. We
summarize the results in Sec. 4.

The notation used in this paper is as follows:  the vectors and matrices 
as a whole are denoted by bold letters. 
The $l_{0}$-(quasi)norm of a vector $\vec{z}$, denoted as
$\norm{\vec{z}}_{0}$ is defined as the number of non-zero elements of
$\vec{z}$; the $l_{1}$-norm of $\vec{z}$ is defined as $\norm{\vec{z}}_{1} =
\sum_{i=1}^{n} |z_{i}|$. The $l_{0}$- and  $l_{1}$- 
norms for a matrix $\mat{X}$ are defined by taking it as 
an vector. The Frobenius norm of a
matrix $\mat{X}$ is defined as $\norm{\mat{X}}_{F} =
\sqrt{\Tr{(\mat{X} \mat{X}^{\dagger})}}$, where $\Tr \mat{M}$ denotes the trace of the 
matrix $\mat{M}$. Finally, the vector hard-thresholding operator $\Theta_{\lambda}(\vec{z})$  is defined 
component-wise as 
\begin{equation} \label{eq:ht}
  \Theta_{\lambda}(z_{i}) =
                  \left\{
                    \begin{array}{ll}
                        z_{i}  & \text{if} \quad |z_{i}| > \lambda; \\
                        0 & \text{otherwise}.
                    \end{array}
                  \right.
\end{equation}

\section{Basic Principle}\label{S:bas}
In radio interferometry a visibility $V_{ij}$ is the instantaneous
correlation between the voltages from two receiver feed elements $F_{i}$ and $F_{j}$. 
Without losing generality, we may assume there are two orthogonal polarizations 
$X$ and $Y$ in each feed. In the Tianlai cylinder case, the feeds are dipoles with 
linear polarization, and we shall call the east-west polarization $X$ and north-south polarization
$Y$.  The interferometer takes four combinations
of the measurements $V_{ij}^{XX}$, $V_{ij}^{XY}$, $V_{ij}^{YX}$ and
$V_{ij}^{YY}$ for each baseline $(i,j)$. In this paper, we deal with
only the non-polarized calibration, i.e., we do the calibration for
only $V_{ij}^{XX}$ and $V_{ij}^{YY}$ independently. For symbolic
simplicity, we omit the $XX$ and $YY$ superscript in the following
discussion. 
With noise, the voltage of element $i$ is
\begin{equation}
F_i = g_i \int d^2 \uvec{n} A_i(\uvec{n}) \mathcal{E(\uvec{n})} 
e^{-2\pi i \uvec{n} \cdot  \vec{u}_i} + n_i
\end{equation}
where $\mathcal{E(\uvec{n})}$ is the electric field of the radio wave coming from direction 
$\uvec{n}$ on the celestial sphere, 
$A_{i}(\uvec{n})$ is the primary beam of feed $i$, and $g_{i}$ is a
direction-independent complex gain factor that calibration seeks to solve, and $n_i$ is the noise in receiver $i$.
Assuming that the signal and noise are uncorrelated, and neglecting the couplings between the feeds,  
the visibility is ideally given by
\begin{eqnarray} \label{eq:vij}
  V_{ij} &\equiv& \langle F_{i} F_{j}^{*} \rangle \nonumber \\
  &=& g_i g_j^* \int d^{2}\uvec{n} \,
  A_{i}(\uvec{n}) A_{j}^{*}(\uvec{n}) e^{-2\pi i \uvec{n} \cdot
    \vec{u}_{ij}} I(\uvec{n}) + \langle n_i n_j^* \rangle, \nonumber
\end{eqnarray}
where $\vec{u}_{ij} = (\vec{r}_{i} - \vec{r}_{j}) / \lambda$ is the
baseline vector between the two feeds in units of wavelength, 
 and $I(\uvec{n})$ is the sky intensity distribution. 
One can substitute the sky model and telescope model into these equations
to solve for the complex gains.

\subsection{Complex Gain as Eigenvectors}

If we have a good knowledge on the 
primary beam responses $A_{i}(\uvec{n})$, the positions $\vec{u}_{i}$, and a
sky model $I(\uvec{n})$, neglecting noise, we can compute the visibilities induced by the
sky model $V_{ij}^{\text{model}}$. The ratio between the observation and data is
\begin{equation} \label{eq:Rij}
  R_{ij} = V_{ij}^{\rm obs} / V_{ij}^{\text{model}} = g_{i} g_{j}^{*}.
\end{equation}
or in matrix form,
\begin{equation} \label{eq:Rgg}
  \mat{R} = \vec{g} \vec{g}^{\dagger},
\end{equation}
where $\vec{g}$ is a vector with its $i$-th element being the gain
$g_{i}$. 

One can simply go for numerical solution of Eq.~(\ref{eq:Rgg}) by putting in the model 
and observed values. However, taking note of the form of Eq.~(\ref{eq:Rgg}), an eigen-analysis 
method presents itself for solution. Specifically, because $\mat{R}$ is a rank-one
matrix, it has only one non-zero eigenvalue in the absence of noise. Note that 
$\mat{R} \vec{g}=  (\vec{g} \vec{g}^\dagger) \vec{g} = \vec{g}
(\vec{g}^\dagger \cdot \vec{g})$ and $\vec{g}^\dagger \cdot \vec{g}=||\vec{g}||^2 =\sum_i |g_i|^2$ is a real number, 
so the (unnormalized) eigenvector of $\mat{R}$ is $\vec{g}$, with eigenvalue $||\vec{g}||^2$.
Thus, in principle the complex gains of the array could be obtained by solving the eigenvalue problem for 
the matrix $\mat{R}$. 

However, noise is present in actual measurement, and the beam response is not precisely known so the computation 
of the model visibility is inaccurate or even impossible, making the solution with
Eq.~(\ref{eq:Rij}) and (\ref{eq:Rgg})  impractical in the general case.
But if there is a strong radio point source with flux $S_{c}$ at direction $\uvec{n}_{0}$ which
dominates over the noise, then 
\begin{eqnarray}
V_{ij}= V_{ij}^{0} +n_i n_j^*, 
\end{eqnarray}
where $n_i, n_j$ are the noise 
from the receivers $i,j$ respectively, and
\begin{eqnarray}
V_{ij}^{0}  &=& S_{c} \, G_{i} G_{j}^{*} ,
 \label{eq:Vps}
\end{eqnarray}
with
\begin{equation} \label{eq:G_i}
G_i = g_i A_i(\uvec{n}_0) e^{-2 \pi i \uvec{n}_{0} \cdot \vec{u}_{i}};
\end{equation}
in matrix form,
\begin{equation} \label{eq:V0ps}
  \mat{V}_{0} = S_{c} \, \vec{G} \vec{G}^{\dagger}. 
\end{equation}
The vector $\vec{G}$ which includes complex gain and beam response is an eigenvector of $\mat{V_{0}}$. 

If noise is present but small compared with the calibrator source and statistically equal in all elements, 
i.e. $\mat{V}=\mat{V}_0 + \mat{N}$, where $\mat{N}=\langle \vec{n} \vec{n}^\dagger \rangle$, the vector
$\vec{G}$ could be obtained by principal component analysis (PCA):  solving the eigenvector of the matrix 
$\mat{V}$, with the eigenvector associated with the largest eigenvalue identified as $\vec{G}$.  This is also the least
square solution of the form $\mat{V}= \vec{g} \vec{g^\dagger}$.
To prove this, introduce a Lagrangian multiplier $\lambda$, and normalize the solution to 
satisfy 
$$g_i=\sqrt{\lambda} v_i,\qquad \sum |v_i|^2=1,$$
define the residual error 
$$\epsilon \equiv \sum_{i,j} (V_{ij}-\lambda v_i \bar{v}_j)^2.$$ 
The least square solution is obtained by $\partial \epsilon/\partial \bar{v}_i=0$, i.e.
\begin{equation}
\sum_j V_{ij} v_j = \lambda v_i.
\end{equation}
which is the eigenvector equation. Note also that adding a constant along the diagonal of the 
matrix does not change the solution, and for a 
unit normalized covariance matrix, setting 
the diagonals to zero does not affect the solution either.

This is the basic idea of calibration with eigenvector analysis. 
The solution obtained as an eigenvector automatically satisfies both
the phase and the amplitude closure relations. This is because the
quantity $g_{i}$ is of the form $g_{i} = |g_{i}| \, e^{i \phi_{i}}$,
from the algebraic identity $(\phi_{i} - \phi_{j}) + (\phi_{j} -
\phi_{k}) + (\phi_{k} - \phi_{i}) = 0$ and $|g_{i}| |g_{j}| |g_{k}|
|g_{l}| = |g_{i}| |g_{k}| |g_{j}| |g_{l}|$ we always have 
\[
    \text{Arg}(g_{i} g_{j}^{*}) + \text{Arg}(g_{j} g_{k}^{*}) + \text{Arg}(g_{k} g_{i}^{*}) = 0,
\]
and 
\[
    |g_{i} g_{j}^{*}| |g_{k} g_{l}^{*}| = |g_{i} g_{k}^{*}| |g_{j} g_{l}^{*}|.
\]

\subsection{Stable Principle Component Analysis}
In the real world, in addition to the calibrator source and noise, there may be 
radio frequency interferences (RFIs), or some data might be missing due to various reasons, e.g. receiver malfunction. 
Even though some of the RFIs and missing data might be removed in preprocessing, some large residues may still
be present and wreck the PCA. In such a case, the observed visibilities can be modeled as 
\begin{equation} \label{eq:Vpsm}
  \mat{V} = \mat{V}_{0} + \mat{S} + \mat{N},
\end{equation}
where $\mat{V}_0 = S_{c} \, \vec{G} \vec{G}^\dagger$
is a rank 1 matrix from the calibrator (strong point source), $\mat{S}$ is a sparse matrix whose elements are
outliers (the un-flagged RFI, abnormal value, etc) which may have
large magnitude, and $\mat{N}$ is a matrix with dense small elements
which represents the noise,  signal of fainter objects in the field
of view, cross-talks and so on, and we assume it has a magnitude
smaller than the non-zero elements of the outliers $\mat{S}$. The
stable principal component analysis (SPCA) method \citep{Zhou2010} may be applied to solve
the problem in this case. In this approach, the observed data matrix
$\mat{X}$ is decomposed as
$ \mat{X}=\mat{L}+\mat{S}+\mat{N} $
where $\mat{L}$ is a matrix of low rank, $\mat{S}$ a sparse matrix
(i.e. only a small fraction of its elements are non-zero), and
$\mat{N}$ is a dense noise matrix. In our case, the SPCA would yield
$\mat{L} = S_{c} \, \vec{G} \vec{G}^\dagger = \mat{V}_0$.

The SPCA decomposition is achieved by solving the following
optimization problem $$\min_{\mat{L}, \mat{S}} \, \frac{1}{2}
\norm{\mat{X} - \mat{L} - \mat{S}}_{F}^{2} + \lambda
\norm{\mat{S}}_{0} $$ subject to $\rank(\mat{L}) \le r$.
This is done with a block coordinate descent strategy:
first take an estimate of outliers $\mat{S}$ and subtract it out  to get the
``cleaned'' data $\mat{C} = \mat{X} - \mat{S}$, and fit $\mat{L}$ based on
$\mat{C}$. Then, we update the outliers $\mat{S}$ by
hard thresholding on the error $\mat{E} = \mat{X} - \mat{L}$. 
That is, iterate the following steps until it converges:
\begin{enumerate}
\item $\mat{L} = \text{SVD}_{r}(\mat{X} - \mat{S})$;
\item $\lambda = \sqrt{2 \log (m n)} \, \text{MAD}(\mat{X} - \mat{L}) \, / \, 0.6745$;
\item $\mat{S} = \Theta_{\sqrt{2} \lambda} (\mat{X} - \mat{L})$.
\end{enumerate}
Here, $\text{SVD}_{r}(\mat{M})$ is the rank-$r$ truncated SVD of the
matrix $\mat{M}$, i.e., the SVD with all small singular values being
truncated to zero except the largest $r$ ones, $\Theta_{\lambda}
(z_{i})$ is the hard-thresholding operator defined in
Eq.~(\ref{eq:ht}), MAD is the median absolute deviation,
$\text{MAD}(\mat{E}) =
\text{med}(|\mat{E} -  \text{med}(\mat{E})|)$ for a real matrix $\mat{E}$,
where $\text{med}$ denotes the median of the sample. The MAD provides a 
robust estimate for the ``standard error'', in the case of 
independent and identically distributed (i.i.d.) real Gaussian variable
$\mat{E} \in \mathbb{R}^{m \times n}$,
\begin{equation} \label{eq:mad}
  \hat{\sigma} = \text{MAD}(\mat{E}) \, / \, 0.6745. 
\end{equation}
For the complex case, 
\begin{equation}
 \text{MAD}(\mat{E}) = \sqrt{\text{MAD}(\Re{\mat{E}})^{2} +
                        \text{MAD}(\Im{\mat{E}})^{2}} 
                   \label{eq:madE}
\end{equation}

Some entries of $\mat{X}$ may not be available, e.g., the data flagged as RFI. 
Let $\Omega \subset \{ (i, j): i = 1, \cdots, m, j = 1,
\cdots, n \}$ be the index set of the available entries of
$\mat{X}$, and $\Omega_{c}$ be its complement. To deal with the missing data,  we
first set $\mat{X}_{ij} = 0$ for $(i, j) \in \Omega_{c}$ while keeping other
values unchanged,  then solve the optimization
problem as before but with the additional constraint, 
\begin{equation} \label{eq:mis}
  (\mat{L} + \mat{S} + \mat{N})_{ij} = 0, \text{ for } (i, j) \in \Omega_{c}.
\end{equation}
If in the solution the values of these elements in the low-rank component
$\mat{L}$ are close to 0, in $\Omega_{c}$ they will introduce only small perturbations to the data,
which would be separated out as small noises and assigned to the matrix $\mat{N}$. On the other hand, 
if the corresponding values of $\mat{L}$ are large, they will be treated as outliers, as long as the support of the
true outliers and the induced outliers are not too large as to cause the algorithm fail.
This would have little effect for the recovery of the low-rank component $\mat{L}$,
which is usually the one in which we are most interested in practice.

Applying the SPCA to Eq.~(\ref{eq:Vpsm}), we solve for
\begin{align} \label{eq:V0S}
  \min_{\mat{V}_0, \mat{S}} \left[ \frac{1}{2} \| \mat{V} - \mat{V}_{0} - \mat{S} \|_F^2 + \lambda \| \mat{S} \|_0 \right] 
  ~~ \text{s.t.} ~~ \rank(\mat{V}_{0}) \le 1.
\end{align}
To solve Eq.~(\ref{eq:V0S}), we need to initialize the outliers $\mat{S}$. The
simplest choice  would be $\mat{S} = 0$, which works in most cases. Alternatively,  we may set
\begin{equation} \label{eq:Sinit}
  \mat{S} = \left\{
        \begin{array}{ll}
            \mat{V} - \text{med}(\mat{V});  & \text{if } |\mat{V} -  \text{med}(\mat{V})| \ge \tau \, \text{MAD}(\mat{V}); \\
            0 ;& \text{otherwise},
        \end{array}
      \right.
\end{equation}
where $\tau$ is a chosen threshold, usually between 3 and 5. 
The motivation for this initialization is that we expect elements of
$\mat{V}_{0}$ are of similar magnitude, so values that are well above the
median would likely be outliers. This initialization helps to make
the algorithm converge faster.

The SPCA decomposition and eigen-analysis calibration method  
only assumed very simple telescope and sky models. It is fairly robust, and the computation
complexity $\propto N$ instead of $\propto N^{2}$, where $N$ is the
number of elements in the array, which make it scalable to arrays
with a very large number of elements.

\subsection{Extension to Polarization}\label{S:fp}
The method described above can also be extended to case of full 
polarization response calibration with polarized points sources. To 
characterize the full polarization states, in addition to the same polarization 
correlations, we should also include the cross-polarization correlations, i.e.  
 $V_{ij}^{XX}$, $V_{ij}^{YY}$, $V_{ij}^{XY}$, $V_{ij}^{YX}$ for linear polarization feeds, or
$V_{ij}^{LL}$, $V_{ij}^{LR}$, $V_{ij}^{RL}$ and $V_{ij}^{RR}$ for 
circular polarization feeds. 

Denote the electric field of the incoming wave in orthogonal polarization components $\vec{p}$,
$\vec{p} = (v^{X}, v^{Y})$ for  linear polarizations or $\vec{p} = (v^{L}, v^{R})$ for circular polarization.
For a system of $N$ antennas/feeds, the $2N$ component array voltage response is given by  
\begin{equation} \label{eq:qv}
  \vec{q} = \mat{G} \vec{p},
\end{equation}
where $G$ is an $N \times 2$ gain matrix. The observed visibilities are (neglecting noise)
\begin{equation} \label{eq:V}
  \mat{V} = \langle \vec{q} \vec{q}^{\dagger} \rangle = \mat{G} \langle \vec{p}
  \vec{p}^{\dagger} \rangle \mat{G}^{\dagger}.
\end{equation}

If there is one dominating point source, the brightness matrix is $\mathcal{B} = \langle \vec{p} \vec{p}^{*} \rangle $, and 
\begin{equation} \label{eq:BXY}
  \mathcal{B} = \begin{pmatrix}
   I + Q & U + i V \\
   U - i V & I - Q
  \end{pmatrix}.
\end{equation}
for the linear polarization, 
\begin{equation} \label{eq:BLR}
  \mathcal{B} = \begin{pmatrix}
   I + V & Q + i U \\
   Q - i U & I - V
  \end{pmatrix}.
\end{equation}
for circular polarization. Substituting Eq.(~\ref{eq:BXY}) or Eq.~(\ref{eq:BLR}) into Eq.~(\ref{eq:V}),
we see that $V$ is a rank 2 matrix. This is valid for a single dominating source, regardless whether it is 
polarized or not.  As in the un-polarized case, we model the imperfections as noise and outliers, 
\begin{equation} \label{eq:Vsn1}
  \mat{V} = \mat{V}_{0} + \mat{S} + \mat{N},
\end{equation}
where we define
\begin{equation} \label{eq:V01}
  \mat{V}_{0} = \mat{G} \mathcal{B} \mat{G}^\dagger,
\end{equation}
which is the visibility generated by the point source, and it's a
rank 2 matrix. As in the un-polarized case, $S$ is a sparse matrix
whose non-zero elements are outliers, and $N$ is a small dense noise
matrix.

The decomposition in Eq.~(\ref{eq:Vsn1}) could also be done
by the same SPCA algorithm,  except the rank of $V_{0}$ is 2
instead of 1. But now the problem is how to solve for the gain $\mat{G}$ from Eq.~(\ref{eq:V01}). It is
not possible to uniquely determine $\mat{G}$ by observing a single calibrator source. To obtain the full solution of 
 the system gain $\mat{G}$, three calibrators with different polarizations are needed. 
 
 Below we shall solve the $XX$ and $YY$ polarizations separately without 
 considering the cross-polarization correlations; this is equivalent to the solution of two unpolarized cases. 
 The calibration of the Tianlai array with full polarization response will be deferred to future studies.

\section{Application to the Tianlai Array}\label{S:cal}
For illustration, we apply the calibration method described above
to the Tianlai Cylinder Array,  which consists of three adjacent north-south oriented
parabolic cylinders. A total of 96 dual-polarization feeds are installed on them,
with 31, 32, and 33 units on the three cylinders which all span the same length of 12.4 meters, so that the 
distance between the feeds are 41.33 cm, 40.00 cm and 38.75 cm respectively. This arrangement  
forms slightly unequal baselines in order to reduce the grating lobes \citep{Zhang:2016miz}. 
The data set used here was collected during
the first light drift scan observation on 27 September, 2016. 
We choose a period of the data when the sky calibration source Cygnus A (Cyg A) 
transits over the array. The transit time is 13:25:46 (UT+0h).
Before calibration, the data is first pre-processed to remove 
known bad channels and strong radio frequency interferences (RFIs). 
The SumThreshold method \cite{Offringa2010} is used for RFI flagging. 
As an example, here we show the result for the frequency channel of $\nu=750 \MHz$.

Due to logistic reasons, the Tianlai array antenna is located a relatively long distance away from the 
station house, where the electronic systems sit. The radio frequency signal from the antenna feed, 
after first being amplified by a low noise amplifier (LNA), is converted to an analog optical signal and transmitted 
via optical fiber (RF over fiber) to the station house. The cable length is about 7 km, and varies slightly as 
the environment temperature changes.  
This necessitate a two step calibration procedure: in the first step,  we use a periodically broadcasted
artificial noise source signal to do a relative phase calibration, so as to compensate the phase 
variations over time induced by the cable delay;  then we perform 
an absolute calibration by using a strong radio source on the sky.

\subsection{Noise Source Calibration}\label{S:nscal}
The signal from the artificial noise calibrator is 
much stronger than the signal from sky, and its broadcasting time is known, so it is easily 
recognized in the data.  The noise source can be viewed as a near-field source. Approximately,  its visibility
can be approximated as
\begin{eqnarray} \label{eq:vnf}
  V_{ij}^{\text{ns}} &\approx& S_n
  \frac{A_{i}(\uvec{n}_i) A_{j}^{*}(\uvec{n}_j)}{\sqrt{\Omega_{i} \Omega_{j}}} \frac{r^{2}}{r_{i}
    r_{j}} e^{-i \vec{k}\cdot (\vec{r}_{i} - \vec{r}_{j})}
  \end{eqnarray}
where  $r_i, r_j$ are the distance between the noise source and the receiving feed $i, j$ respectively.
However, even if the pre-factor in Eq.~(\ref{eq:vnf} is not exact, the phase factor $e^{-i \vec{k}\cdot (\vec{r}_{i} - \vec{r}_{j})}$ 
would still be right. 
The noise source is switched on and off periodically, and the time averaged visibility data obtained are
\begin{align*}
  V_{ij}^{\text{on}} &= G_{ij} (V_{ij}^{\text{sky}} + V_{ij}^{\text{ns}} + n_{ij}) \\
  V_{ij}^{\text{off}} &= G_{ij} (V_{ij}^{\text{sky}} + n_{ij}),
\end{align*}
so we have
\begin{align*}
  V_{ij}^{\text{on}} - V_{ij}^{\text{off}} &= G_{ij} V_{ij}^{\text{ns}} + \delta n_{ij}\\
                                   &\approx C |G_{ij}| e^{-i k \Delta L_{ij}} e^{-i k (r_{i} - r_{j})} ,                                   
\end{align*}
where $\delta n_{ij}$  is the difference of the random noise.  
As the calibrator signal is much stronger, we can neglect the noise. 
$\Delta L_{ij}$ is the equivalent instrument delay 
difference between the channels $i$ and $j$, which are mostly due to the variation in the 
cable length for the two channels. The phase of it is
\begin{equation} \label{eq:nphs}
\phi_{ij}=  \text{Arg}(V_{ij}^{\text{on}} - V_{ij}^{\text{off}}) = k \Delta L_{ij} + \mathrm{const.},
\end{equation}
where we have used the fact that the relative position of the noise
source and the receiving feeds are fixed and  the
coefficient factor $C$ is stable.

\begin{figure}[ht]
  \includegraphics[width=0.4\textwidth]{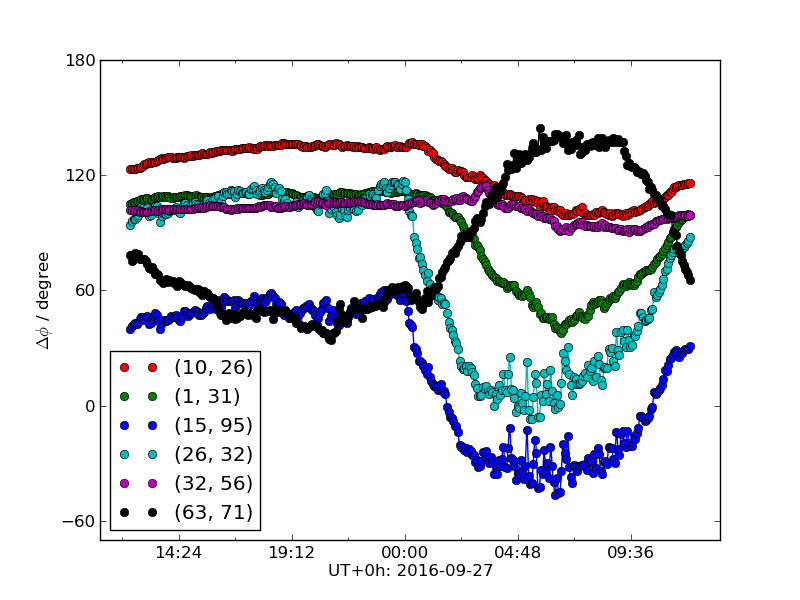}\\
  \includegraphics[width=0.4\textwidth]{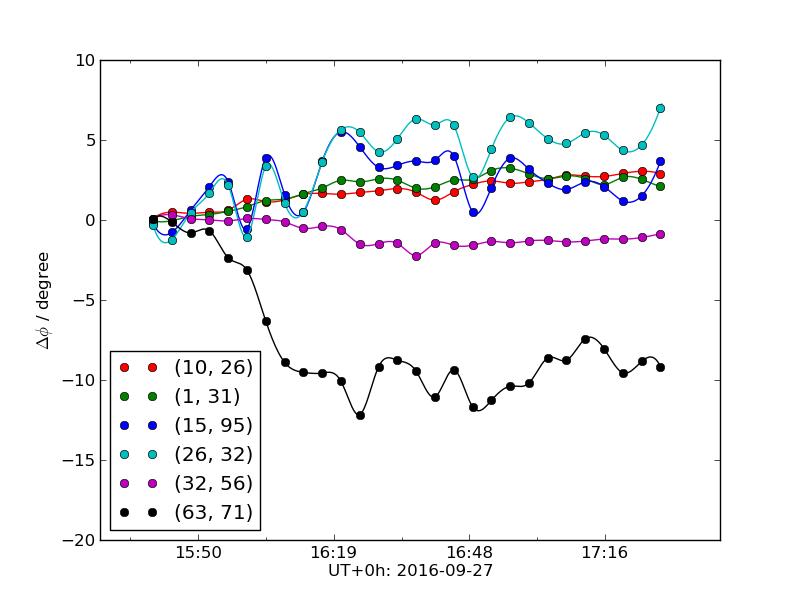}
  \caption{The phase change of some baselines. The $XX$ polarization of the pairs $(10, 26)$, $(15,
    95)$, $(1, 31)$, $(26, 32)$, $(32, 56)$ and $(63, 71)$ are plotted. Top: phase changes during one day; Bottom: phase changes
    during two hours in the night. The results for $YY$ signal are similar. 
    .}
  \label{fig:phsc}
\end{figure}

For most receivers the calibrator is in the side lobe of the beam, so 
the observed amplitude is not very stable. We therefore only use it 
to correct for the phase change. An example of this phase 
change is shown in Fig.\ref{fig:phsc} for the $XX$ linear polarization of several baselines (the baselines are marked by the pair of 
receiver number). We see the phases change smoothly for a small amount during the night, varying a few degrees  for most baselines. 
However, the changes are more significant and rapid during day time, when the
temperatures varies more significantly, which affects the length of the optical fibers.

We compensate for the relative phase change due to $\Delta L$ in the relative phase calibrated visibility:
\begin{equation} \label{eq:comp}
  V_{ij}^{\text{rel-cal}} = e^{-i \phi_{ij}}  V_{ij}.
\end{equation}
After this step, the phase variation over time in the observed
visibility is removed, but there is still an unknown constant phase factor
to be determined in $V_{ij}^{\text{rel-cal}}$, which must be determined from 
absolute calibration using sky source. 

\subsection{Sky Source Calibration}\label{S:pscal}
After compensating for the relative phase changes, we use the calibrator on-off
data for sky calibration. In this first stage calibration, we use the strongest radio point sources
during its meridian transit as the calibrator. Cyg A is an excellent source for such 
purpose, as its position is very close to the zenith of the array. We also used several other strong sources,
such as the Cassiopeia A (Cas A), Taurus A (Tau A), Virgo A (Vir A).
In this first stage calibration, we calibrate the $XX$ and $YY$
linear polarizations separately; the full-polarization calibration
will be deferred to future works.

\begin{figure}[htbp]
\centering
\includegraphics[trim=30 10 30 10,clip,width=0.4\textwidth]{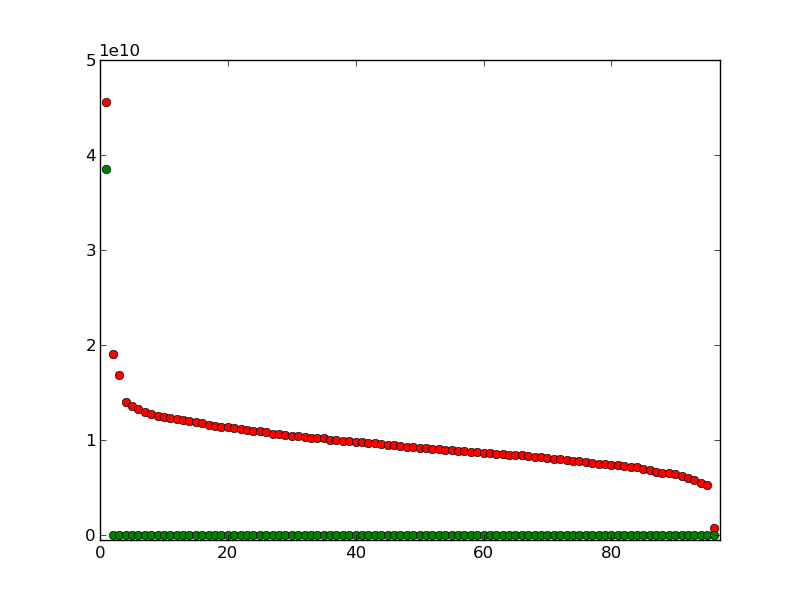}\\
\includegraphics[trim=30 10 30 10,clip,width=0.4\textwidth]{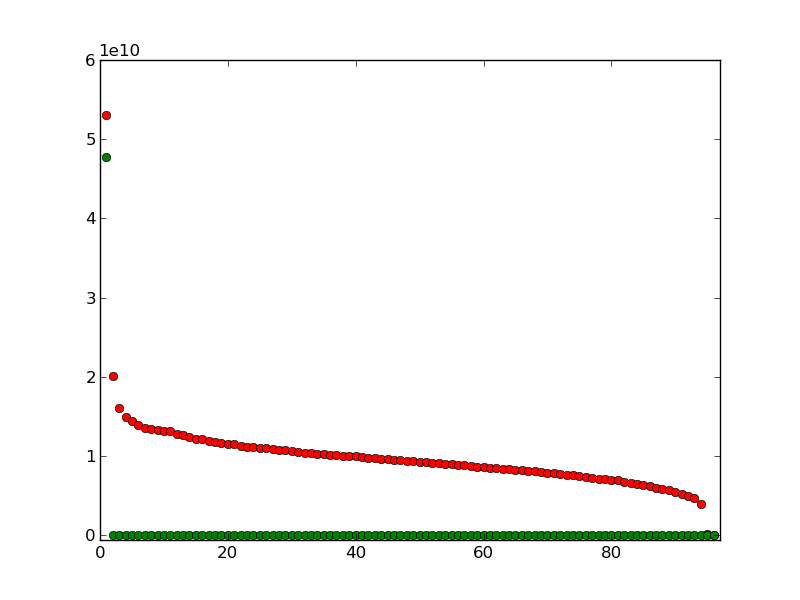}
\caption{The eigenvalues of $\mat{V}_{0}$ (green) and $\mat{V}$
  (red). The top and bottom panels are for the $XX$ (East-West)
  polarization and $YY$ (North-South)
polarization respectively. 
}
\label{fig:eigvalue}
\end{figure}

As shown by the discussions above, in the ideal case of a single dominating point
source we should have only one non-zero eigenvalue in the visibility $\mat{V}$. In
Fig.~\ref{fig:eigvalue}, we show the eigenvalues for the observed visibility data 
$\mat{V}$ (red points). Actually, besides the largest one, there are also a few 
other sizable eigenvalues, which are perhaps due to the effect of outliers or noises. In contrast, 
 we also show the matrix $\mat{V}_{0}$ obtained by the SPCA decomposition  (green points) on the same plot.
In this case there is a single large value, and the remaining eigenvalues are all very small;
the SPCA decomposition helps to separate out the components and get better calibration precision. 

\begin{figure}[htbp]
\centering
\includegraphics[trim=30 10 30 10,clip,width=0.4\textwidth]{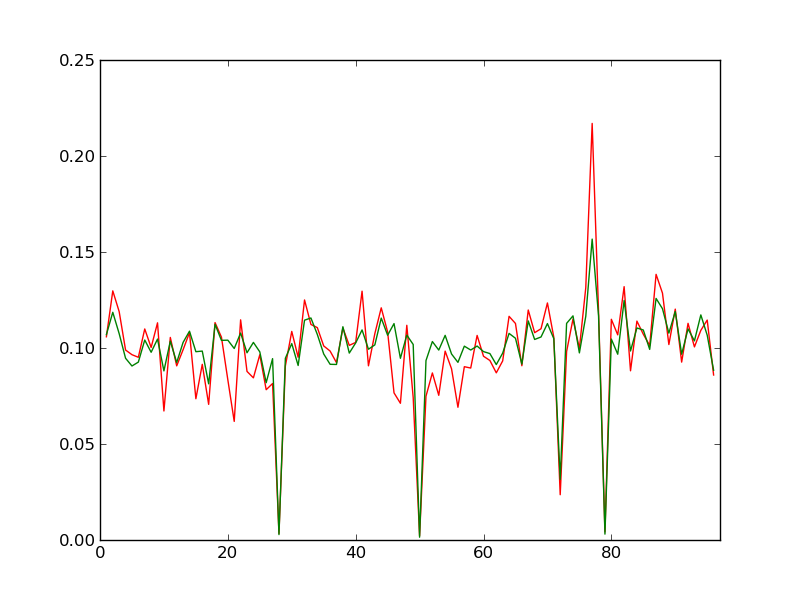}\\
\includegraphics[trim=30 10 30 10,clip,width=0.4\textwidth]{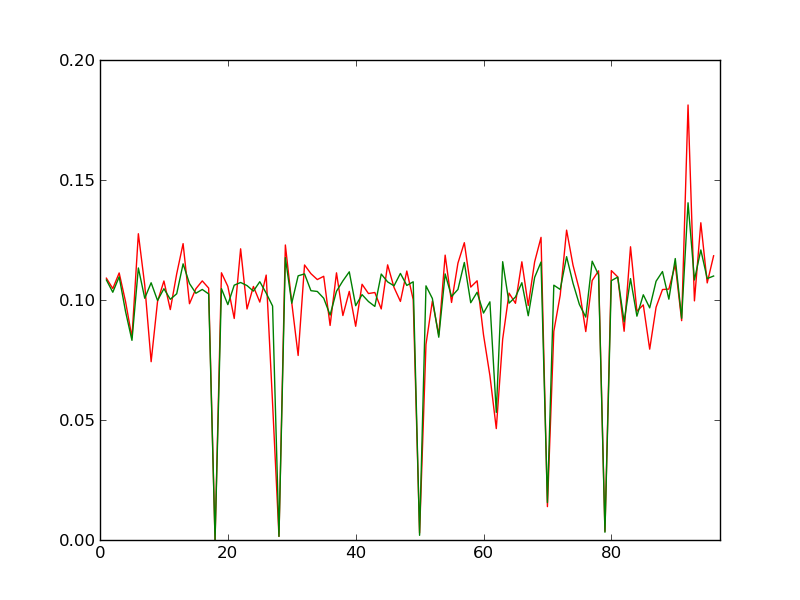}
\caption{Magnitude of the eigenvector corresponding to the largest eigenvalue of
  $\mat{V}_{0}$ (green) and $\mat{V}$ (red). 
  The top and bottom panels are for the $XX$ (East-West) polarization and $YY$ (North-South) polarization
respectively. 
}
\label{fig:eigvector}
\end{figure}

In Fig.~\ref{fig:eigvector} we plot the magnitude of the eigenvector
corresponding to the largest eigenvalue of $\mat{V}$ (red) and $\mat{V}_{0}$ (green). 
The  eigenvector is taken as the solution of
$\vec{G}$, from which the gain $g_{i}$ of each receiver unit is
obtained. Note that for $\mat{V}_0$, several gain values are nearly zero, which 
are due to malfunctioning hardware. The SPCA automatically separated these out as outliers.
The other gain values are slightly affected, but generally the magnitude are comparable with each 
other.  So we see the presence of large outliers biases the principal components estimation, but the SPCA
may help remove these outliers.

\begin{figure*}[htbp]
  \centering
  \subfloat[$\mat{V}$\label{fig:Vx}]{\includegraphics[trim=50 35 40 24,clip,width=0.45\textwidth]{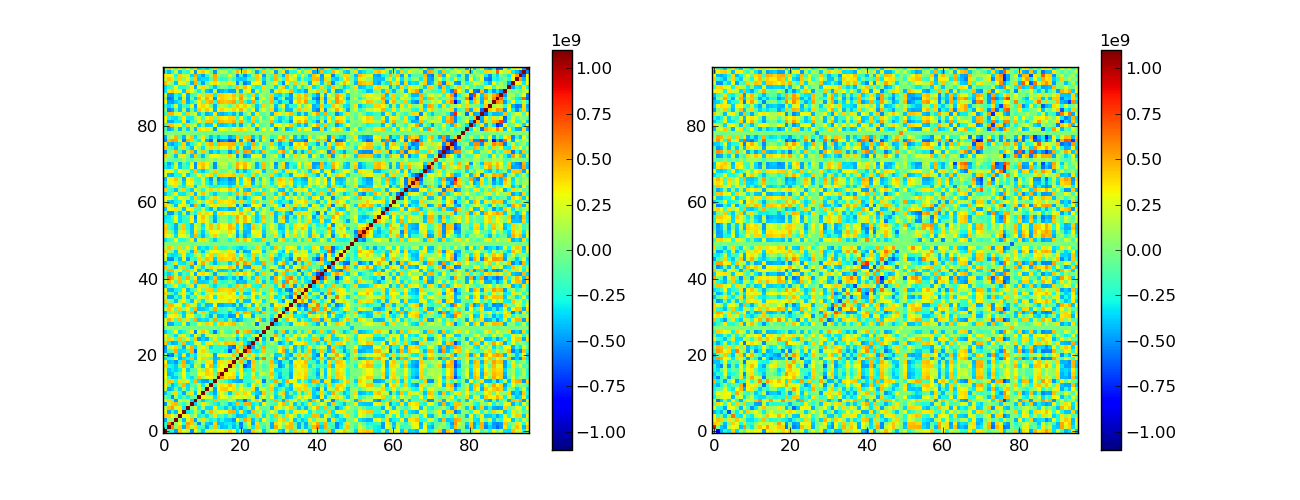}} 
  \subfloat[$\mat{V}_{0}$\label{fig:V0x}]{\includegraphics[trim=50 35 40 24,clip,width=0.45\textwidth]{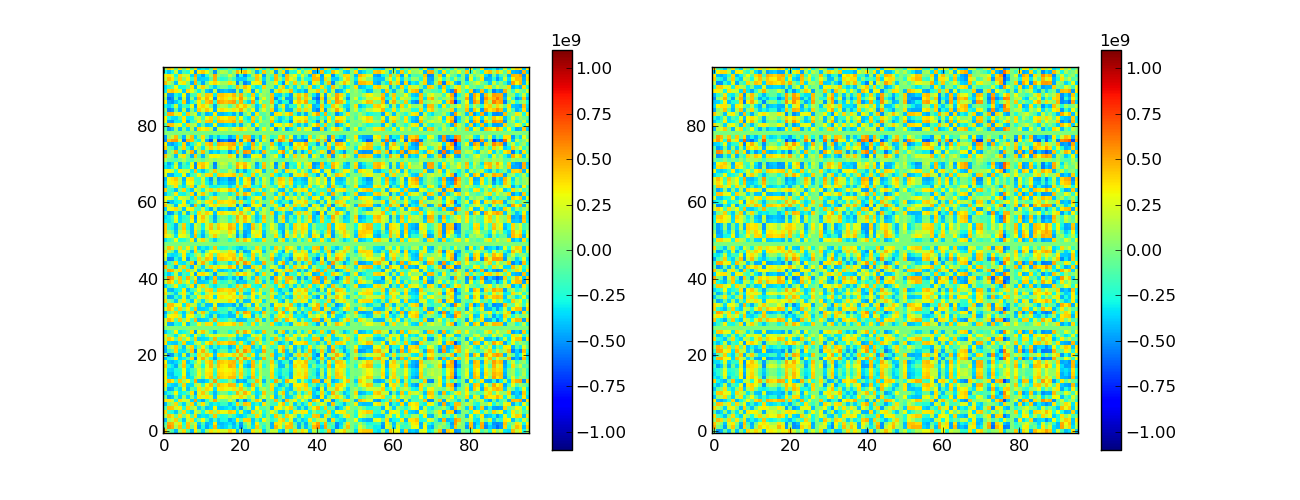}} \\
  \subfloat[$\mat{S}$\label{fig:Sx}]{\includegraphics[trim=50 35 40 24,clip,width=0.45\textwidth]{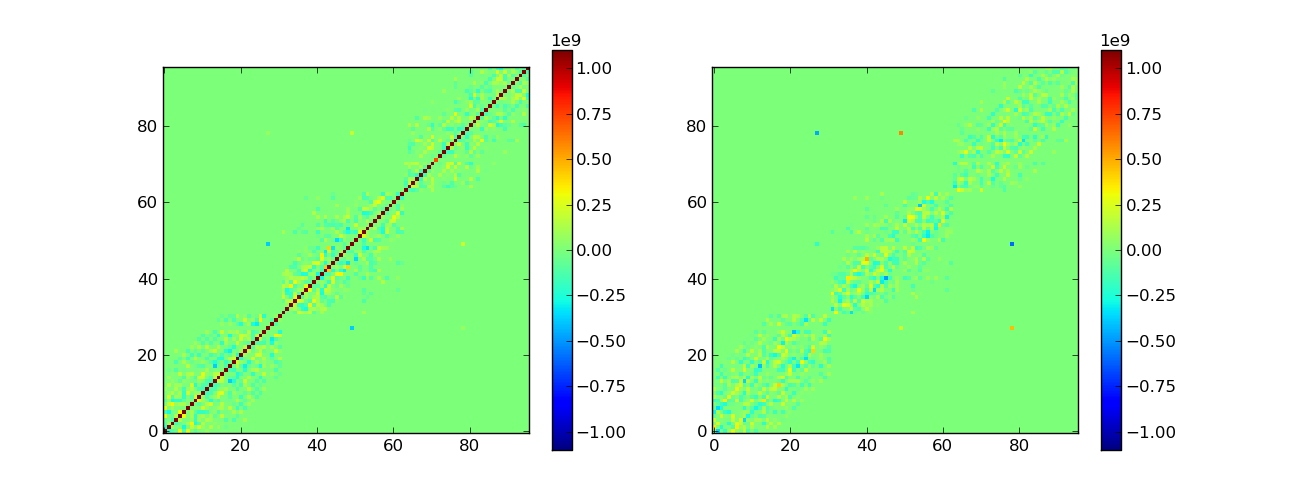}} 
  \subfloat[$\mat{N}$\label{fig:Nx}]{\includegraphics[trim=50 35 40 24,clip,width=0.45\textwidth]{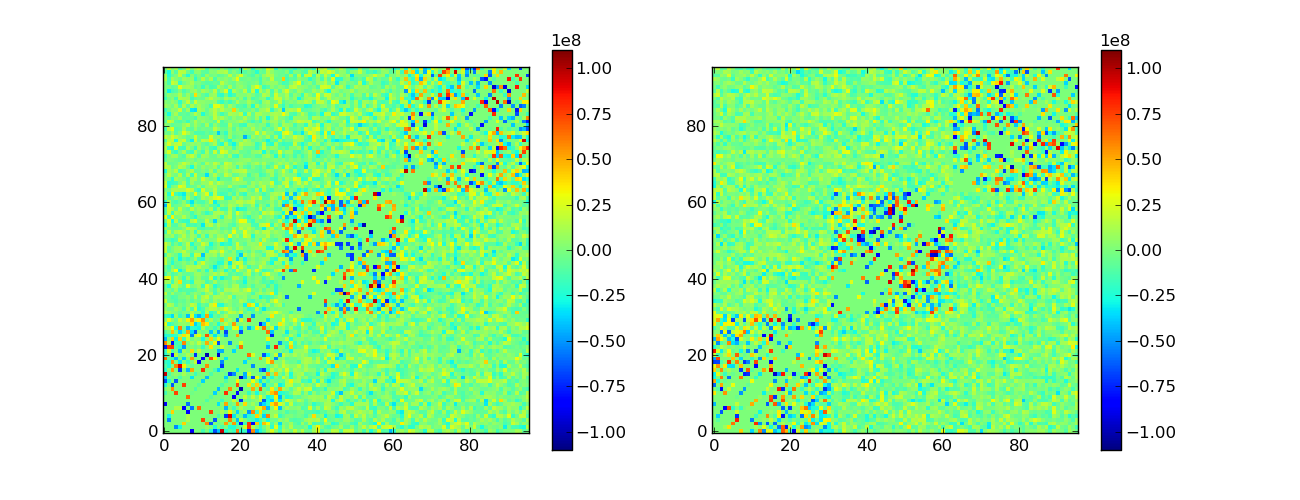}} \\
   \subfloat[$\mat{V}$\label{fig:Vy}]{\includegraphics[trim=50 35 40 24,clip,width=0.45\textwidth]{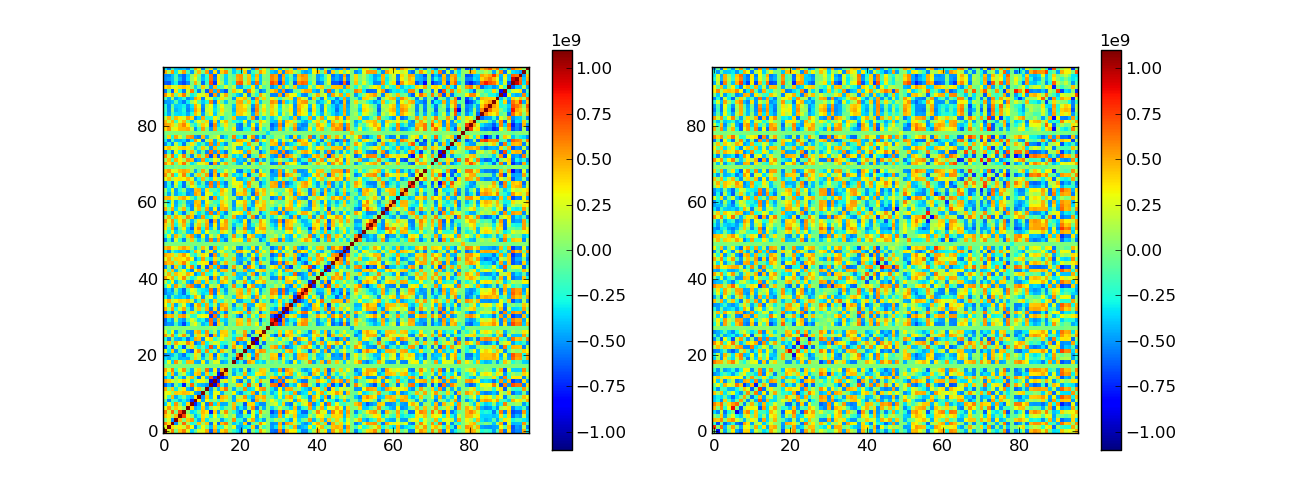}} 
  \subfloat[$\mat{V}_{0}$\label{fig:V0y}]{\includegraphics[trim=50 35 40 24,clip,width=0.45\textwidth]{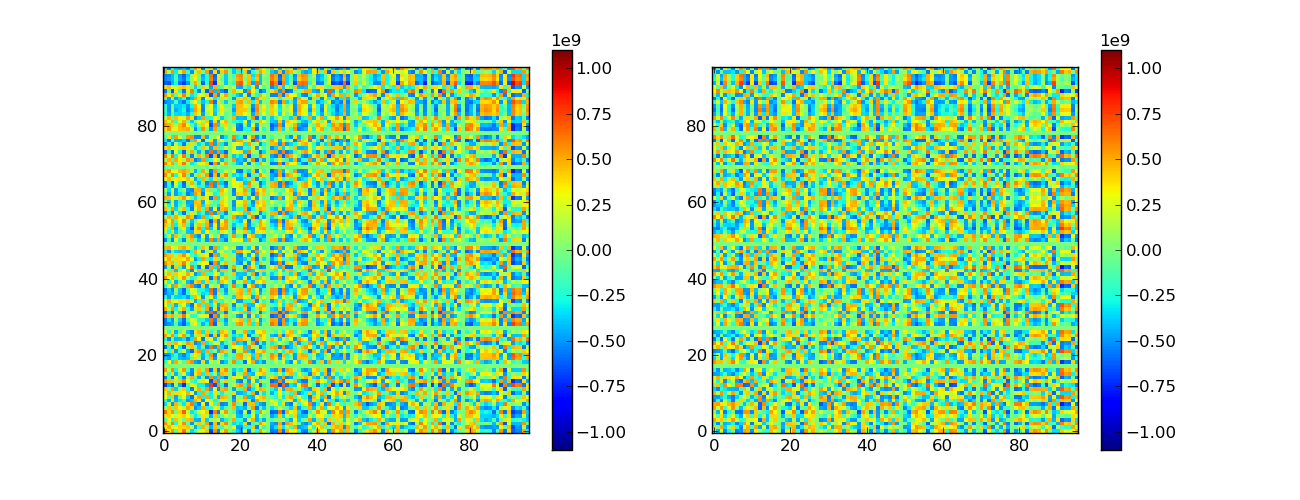}} \\
  \subfloat[$\mat{S}$\label{fig:Sy}]{\includegraphics[trim=50 35 40 24,clip,width=0.45\textwidth]{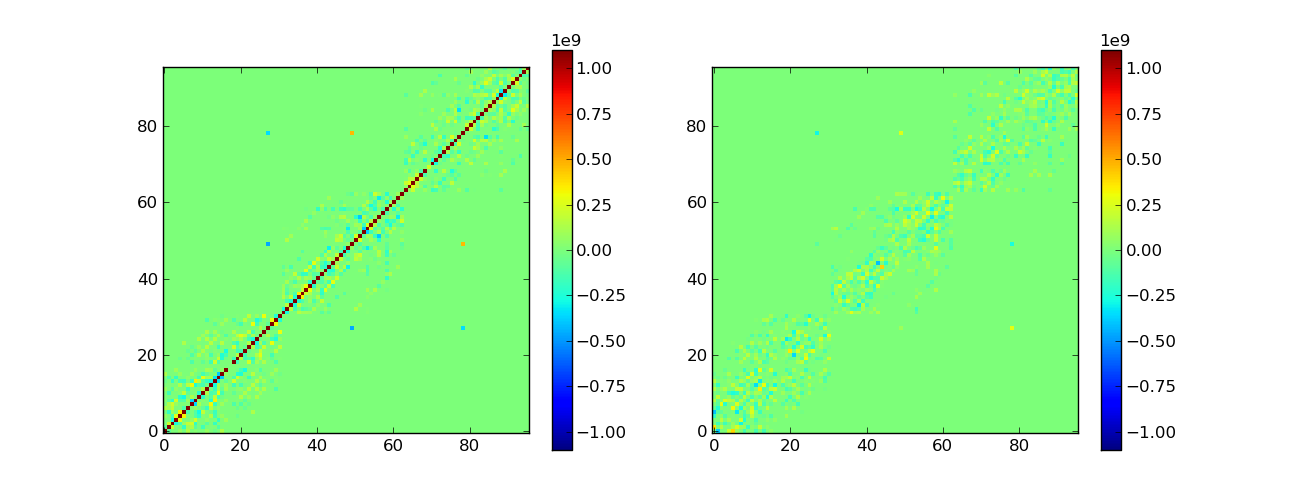}} 
  \subfloat[$\mat{N}$\label{fig:Ny}]{\includegraphics[trim=50 35 40 24,clip,width=0.45\textwidth]{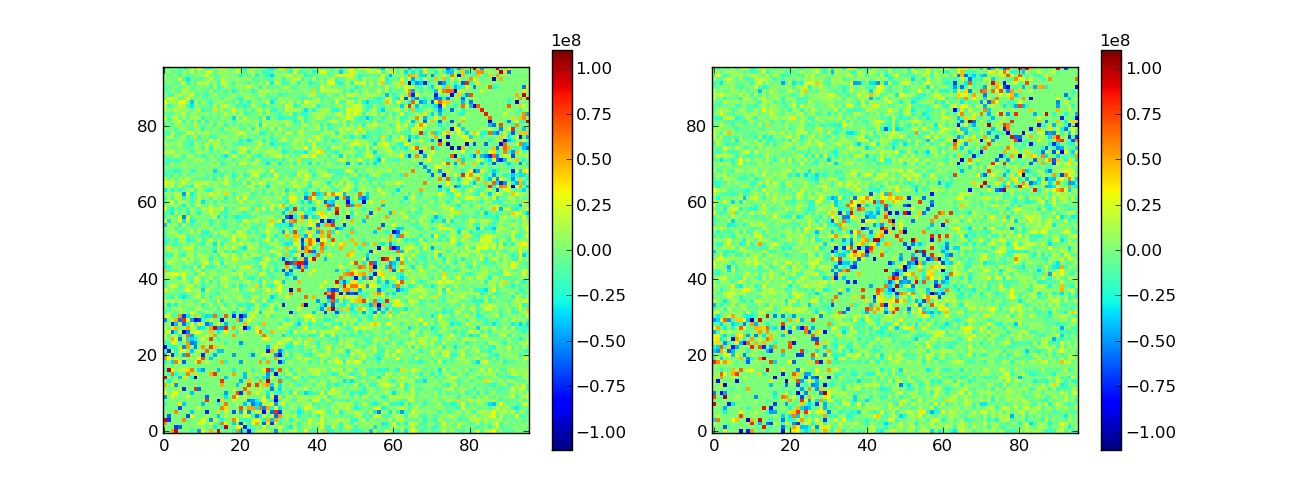}} 
  \caption{SPCA Decomposition of the $XX$ (top, a,b,c,d) and $YY$(bottom,e,f,g,h) visibilities for the Cyg A transit. 
  For each sub-figure, left is the  real part of the data and right is its imaginary part. 
}
  \label{fig:decomp}
\end{figure*}

We show the  SPCA decomposition of the observed visibility 
in Fig.~\ref{fig:decomp} for the Cyg A transit. The three components are
successfully separated.  Although the
auto-correlation (the main diagonal of the $\mat{V}$ matrix) of each feed
is high, it does not appear in the recovered rank-one matrix
$\mat{V}_{0}$. The auto-correlation is  dominated by noise, but amazingly,  
the much smaller visibility induced by the sky source 
is extracted from the data. Also we note that in the recovered $\mat{V}_{0}$, 
there are several apparently symmetric horizontal/vertical strips that have value of 0. 
We have checked that they correspond to the bad feeds, which are automatically detected in this decomposition. 
The outliers are picked out and put in the sparse $\mat{S}$ as
expected. Though we call $\mat{S}$ the outliers matrix, not all of its
non-zero elements are outliers. The high noise in the auto-correlations and 
short baselines also come under  $\mat{S}$ in this classification. 
For the same reason, elements in $\mat{N}$ are not all
pure random noise, and we can see some obvious patterns in it. Three squares
along the main diagonal are formed by the correlations/cross-talks between feeds along 
 the same cylinder.
 
\begin{figure}[htbp]
\centering
\includegraphics[width=0.48\textwidth]{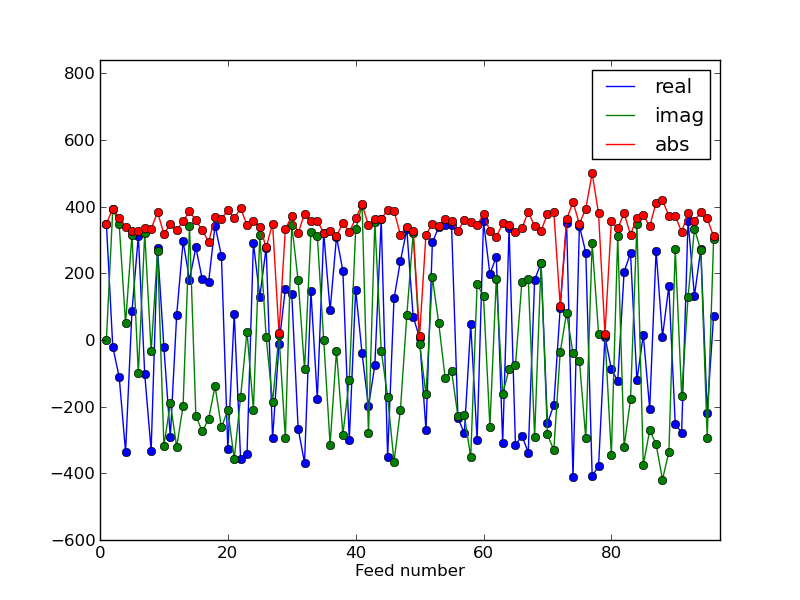}\\
\includegraphics[width=0.48\textwidth]{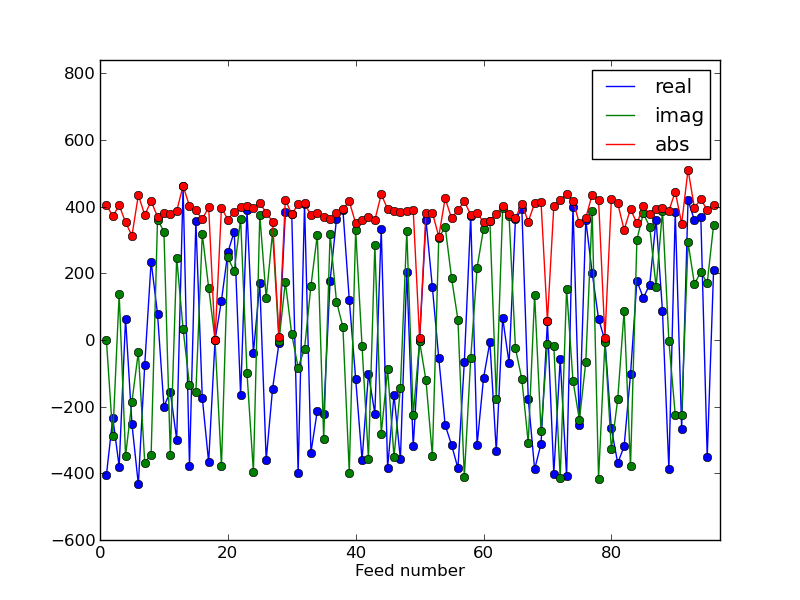}
  \caption{The snapshot of solved $\vec{G}$ for $XX$(top panel) and $YY$(bottom panel) polarization at the transit
    time of Cyg A. }
  \label{fig:G}
\end{figure}

In Fig.~\ref{fig:G} we show a snapshot of solved $\vec{G}$ at the
transit time of Cyg A obtained from the SPCA analysis. 
Both the real and imaginary parts as well as the amplitude of $G_i$ are shown. We see that the phase of $\vec{G}$ are randomly distributed, but most feeds has a typical $|G_{i}|$ value of $300\sim 400$ (digital output, arbitrary units). 
But a few  feeds have small gain amplitude, $|G_{i}| \approx 0$; these are the malfunction ones.

If the beam response $A_i(\hat{\boldsymbol{n}})$ 
and the positions of the antenna/feed $\vec{u}_{i}$ are accurately known, we can solve the gain
$g_{i}$ for each feed $i$ from $G_{i}$. But the beam response of the
Tianlai cylinder array has not been measured before. While a beam model was computed 
with electromagnetic field simulation \citep{2017JAI.....650003C},
it is based on the ideal model, while the actual construction could be different. 
Here we fit the beam profile from the observed data.

From Eq.~(\ref{eq:G_i}), the normalized $G_i$ is given by
\begin{equation} \label{eq:GoG}
  \hat{G}_{i} \equiv \frac{G_{i}}{|G_{i}|} = \hat{g}_{i} E_{i},
\end{equation}
where $\hat{g}_{i} \equiv g_{i}/|g_{i}|$ and and  $E_{i} = e^{-2 \pi i \hat{\boldsymbol{n}}_0 \cdot \vec{\boldsymbol{u}}_i}$.
 We used the fact that the beam profile
$A_i(\hat{\boldsymbol{n}}_0)$ is real. 
$E_i$ varies as the calibration source transits over the beam. Assuming that the receiver is 
stable and its complex gain $g_{i}$ a constant during this period, we may fit $\hat{g}_{i}$ with the observational data.
This determines the phase of the gain $g_{i}$. For the amplitude $|g_{i}|$,
\begin{equation} \label{eq:Giabs}
  | G_{i} | = | g_{i} | \, A_i(\hat{\boldsymbol{n}}_0),
\end{equation}
We see $|g_i|$ is degenerate with the normalization of 
beam response $A_i(\hat{\boldsymbol{n}}_0)$. We may choose a normalization, e.g., 
take $A_i(\hat{\boldsymbol{z}}) = 1$ in the direction of zenith. In fact, 
it happens that the declination of Cyg A ($40^\circ 44^\prime 02^{\prime\prime}$) is close to the latitude of the 
Tianlai site ($44^\circ 09^\prime 08^{\prime\prime}$), so it crosses near the Zenith during its transit. 
The cylinder array beam response is a narrow strip along the
north-south direction and it varies slowly near the zenith, so as a first approximation 
we can normalize $A_i \approx 1$ at the direction of 
Cyg A when it transits over the array.

\begin{figure}[htbp]
  \centering
  \includegraphics[width=0.45\textwidth]{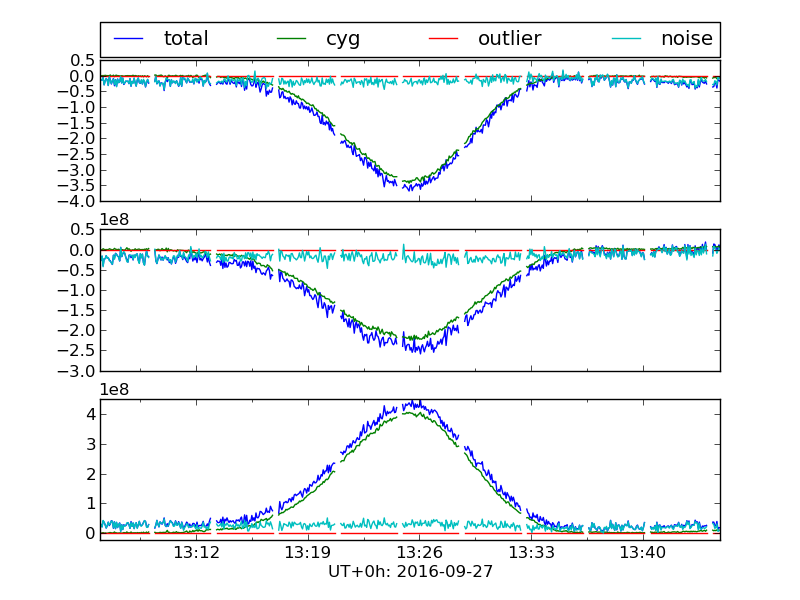}\\
  \includegraphics[width=0.45\textwidth]{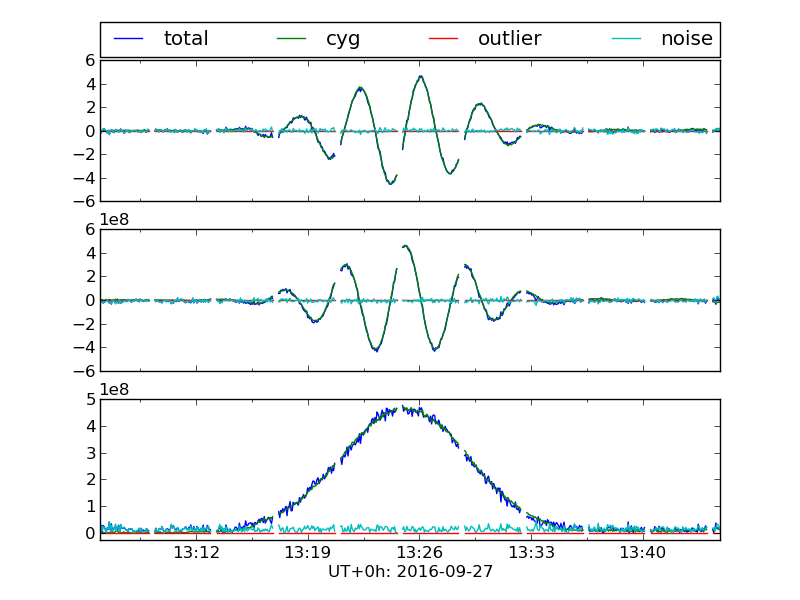}
  \caption{Samples of SPCA decomposition result with nearly zero outliers during Cyg A transit.
    Top: a north-south short baseline $(1, 31)$ XX polarization; Bottom:  an east-west long baseline
    $(15, 80)$ XX polarization. 
    The three subpanels in each plot are, from top to bottom, the real part, imaginary part and the magnitude of the 
    visibility.}
  \label{fig:Vdec}
\end{figure}

\begin{figure}[htbp]
  \centering
  \includegraphics[width=0.45\textwidth]{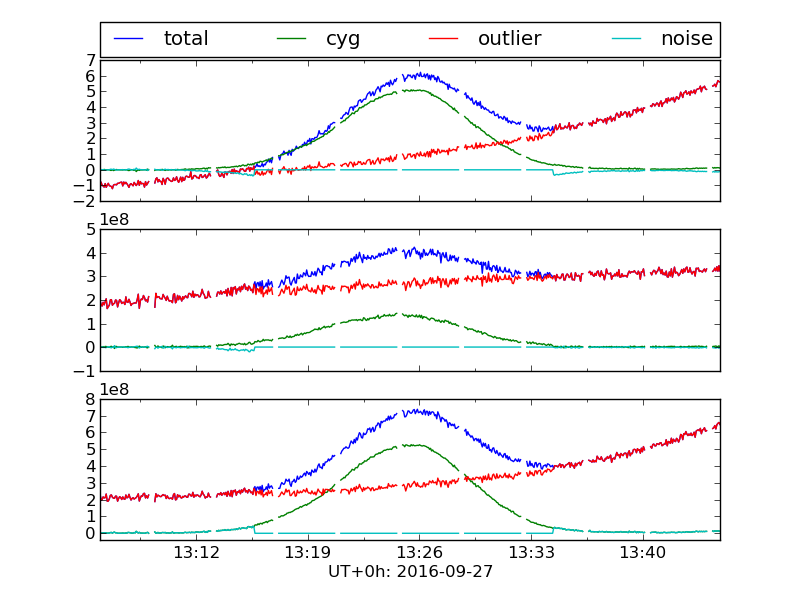}\\
  \includegraphics[width=0.45\textwidth]{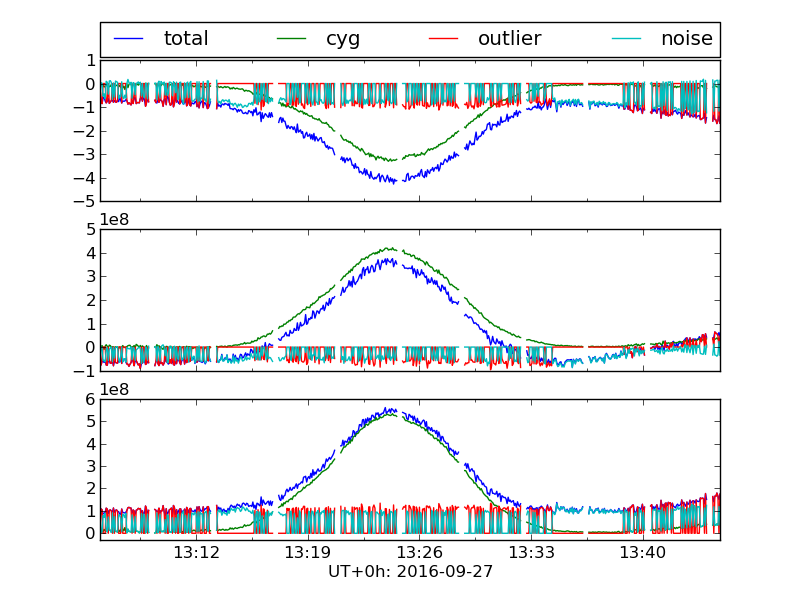}
  \caption{Samples of SPCA decomposition result with non-zero outliers during Cyg A transit.
  Similar to Fig.~\ref{fig:Vdec}, but for two different baselines which exhibit the non-zero outliers. 
  Top: baseline $(2, 3)$ XX polarization; 
  Bottom: baseline $(32, 40)$ XX polarization. }
  \label{fig:Vdec1}
\end{figure}

In Fig.~\ref{fig:Vdec}, we show the total sky visibility and the various components (the Cyg A, the outlier
and the noise) obtained by decomposition for two baselines during a period of 40 minutes centered
at Cyg A's transit time. The two baselines shown are for the element pair
$(1, 13)$ (short, north-south direction)  and the element pair $(15, 80)$(long, nearly east-west) XX polarization.
We have removed the part when the artificial noise calibrator was on, so the curves are broken at the time of their broadcasting. 
As expected, for the Cyg A (marked "cyg" in the figure) component ($\mat{V}_0$),  the 
NS baseline show a general profile of the primary beam, while the 
EW baseline shows interferometer fringes with primary beam as the envelope. 
The outlier and noise components are small for these two baselines during the Cyg A transit. 

The outlier matrix is a sparse one, for most baselines it  is small, as in the last figure, but 
occasionally the decomposition yields non-zero outlier components; two examples are shown in Fig.\ref{fig:Vdec1}.  
These are more frequently seen in the visibility of the short baselines,  
which perhaps have higher noise levels due to cross-interference. 
In the top panel of Fig.~\ref{fig:Vdec1}, the outlier component is much greater than the threshold and so
varies smoothly during the observation. In the bottom panel, as the level of the ``outlier" component  
is close to the threshold $\lambda =\sqrt{2\log{(mn)}} \, \hat{\sigma}$, there is some degeneracy of the two components and 
we can see the ``mixing'' or rapid switching of the two during the observation. This however does not
affect the calibration which uses only the point source component which is still stable.

\begin{figure}[htbp]
  \centering
  \includegraphics[trim=30 10 30 10,clip,width=0.5\textwidth]{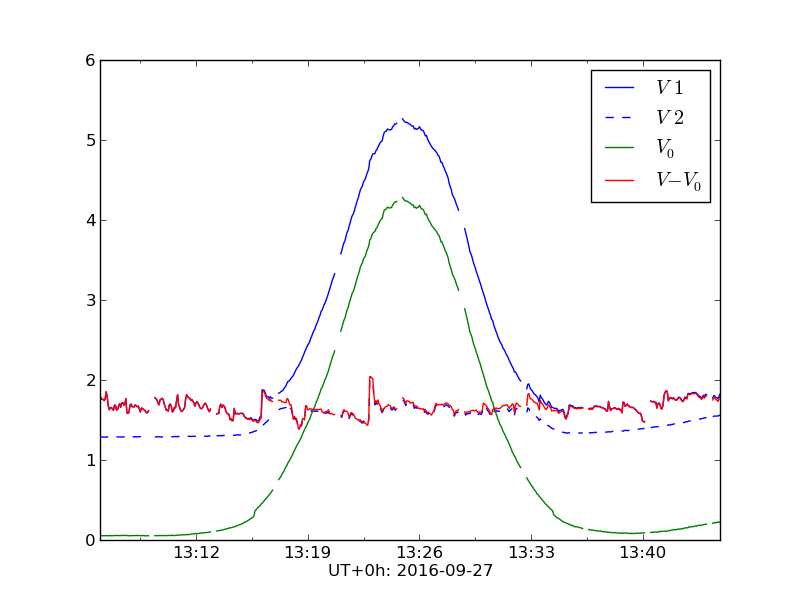}
  \caption{the variation of
the largest eigenvalue of $\mat{V}$, $\mat{V}_{0}$ and $\mat{V}
-\mat{V}_{0}$.}
  \label{fig:eigvs}
\end{figure}

We see the signal of the point source Cyg A is dominant
in about half an hour. When its signal dominates, the SPCA
algorithm can successfully extract it from the observed visibility,
but the algorithm fails when its strength drops to the level of noise. 
When the algorithm fails, the solution of the low-rank component is somewhat unstable. 
The relative strength between the signal and the noise level can be roughly quantified by the 
ratio between the largest eigenvalues of $\mat{V}_{0}$ and $\mat{V} - \mat{V}_{0}$.
In Fig.~\ref{fig:eigvs}, we show the largest and second largest eigenvalues of the matrix $\mat{V}$
(marked as $V_1, V_2$) in solid and dashed blue curves respectively, 
and the largest eigenvalue of the SPCA component
$\mat{V}_{0}$ in green curve, as well as the eigenvalue of the matrix $\mat{V} -\mat{V}_{0}$  in red curve 
during the transit.  The largest eigenvalue of $\mat{V}$ is significantly larger than the second from 13:19 UT to 13:32 UT. 
At the same time, the largest eigenvalue of $\mat{V}_0$ is significantly larger than the largest 
eigenvalue of the remaining components $\mat{V}-\mat{V}_0$, showing the dominance of the calibrator signal.  
Beyond this time interval, the eigenvalues of $\mat{V}$ become comparable with each other, and 
the largest eigenvalue of $\mat{V}_{0}$ drops below that of $\mat{V} -\mat{V}_{0}$.  The algorithm begins to fail to 
extract the low-rank signal component as the signal strength drops. 
We can truncate the algorithm here. However, in practice we find that the algorithm can go a much longer way 
until the largest eigenvalue of $\mat{V}_{0}$ drops below a factor $c < 1$ of that of $\mat{V} -\mat{V}_{0}$. To make
the computation run smoothly, we make the following remedy: when the largest non-zero eigenvalue of the
solved low-rank component $\mat{V}_{0}$ falls below a factor $c$ (we take $c=0.2$) of the largest
eigenvalue of the residual matrix $\mat{V} - \mat{V}_{0}$, we take $\mat{V}_{0}^{\rm eff} = \text{SVD}_{1}{(\mat{V}-\mat{V}_0})$. 
This makes the algorithm works more 
smoothly during a run,  but note that this solution is no longer the true dominating low-rank components (i.e.,
the visibility matrix of the point source). 

\begin{figure}[htbp]
  \centering
  \subfloat[XX(EW) polarization\label{fig:ewg}]{\includegraphics[trim=30 10 30 10,clip,width=0.4\textwidth]{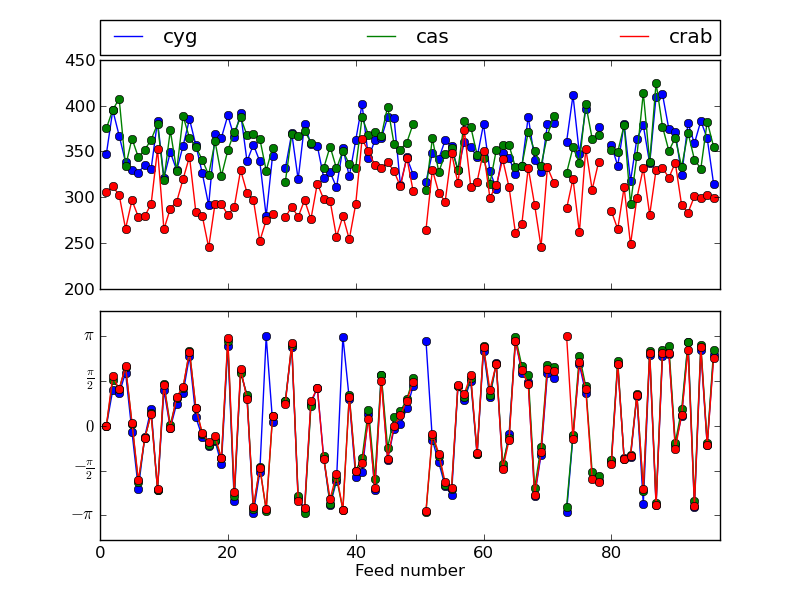}}\\
  \subfloat[YY(NS) polarization\label{fig:nsg}]{\includegraphics[trim=30 10 30 10,clip,width=0.4\textwidth]{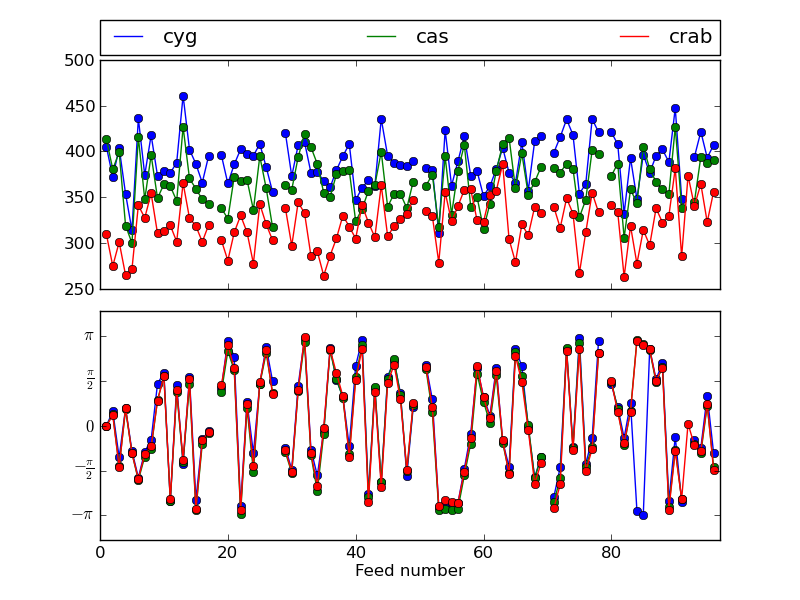}} 
  \caption{The complex gains of (a) The XX(East-West) polarization and (b) YY (North-South) polarization. For each polarization, 
  the amplitude (top) and phase (bottom) are shown for the three different calibrators:  
  Cyg A (cyg), Cas A (cas), and Tau A (crab). }
  \label{fig:gap}
\end{figure}

To check the precision of the calibration, we compare the gains obtained for several
strong point sources with different transit times, including the Cyg A, the Cas A and the
Tau A. They all transit over the array during the night in that observation, over
a time span of about 9 hours. The result is shown in Fig.~\ref{fig:gap}.
We see the complex gain solutions obtained for the three calibrator sources
are highly consistent with each other, especially in their phases. 
The amplitudes of the gains have some differences, but note that in the
approach described above, in each case the beam  $A_i(\hat{\boldsymbol{n}}_0)$ is
normalized to 1 at the peak of the transit, but  the
three calibrators are actually located at different declinations, so part of this
difference may come from the north-south beam profile. 

\subsection{Redundant Baselines}

The redundant baselines of an interferometer array are baselines with the 
same direction and length but formed by different pairs of receivers. Theoretically, 
the redundant baselines should all have identical outputs, so they provide a good 
check on the calibration. The difference in their output reflects the non-uniformity of the 
system.

\begin{figure}[htbp]
\centering
\includegraphics[trim=10 10 30 10,clip,width=0.4\textwidth]{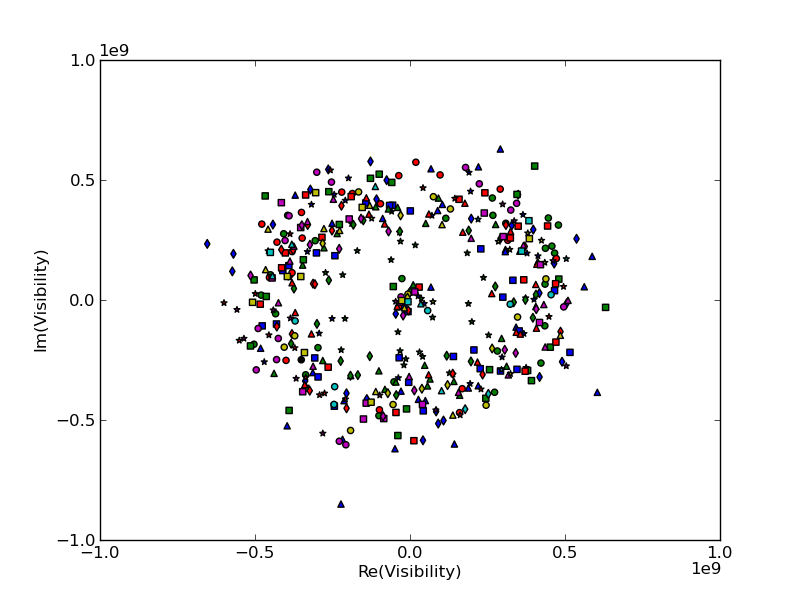}\\
\includegraphics[trim=10 10 30 10,clip,width=0.4\textwidth]{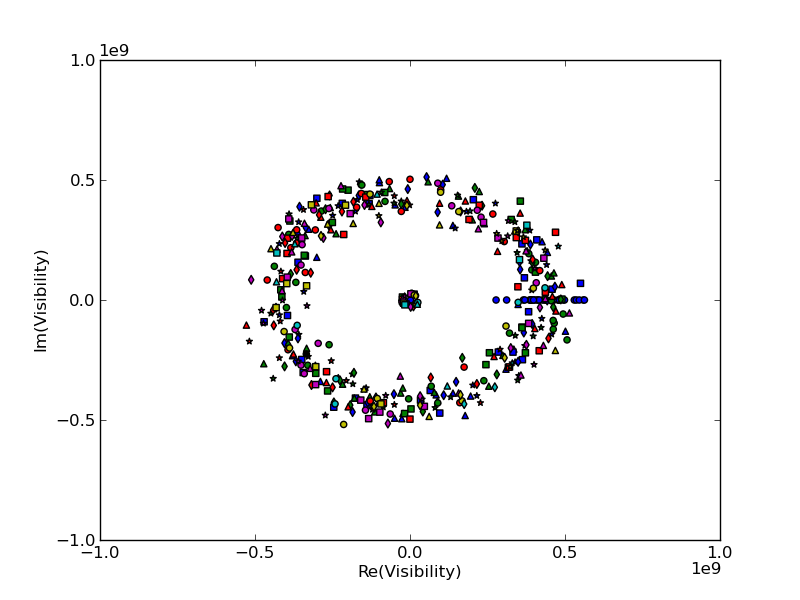}\\
\includegraphics[trim=10 10 30 10,clip,width=0.4\textwidth]{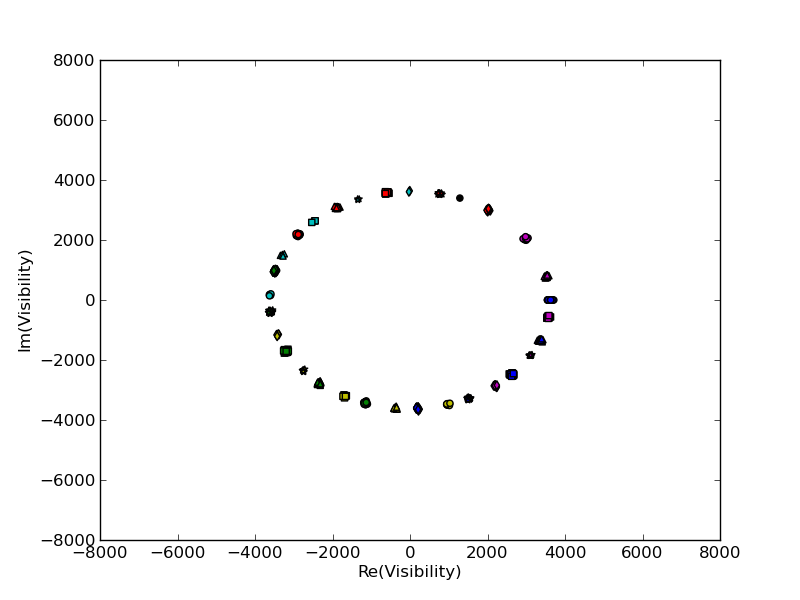}
\caption{The redundant baseline visibilities for one of the cylinder. Top: uncalibrated visibilities; Middle: 
the uncalibrated visibility component $V_0$; Bottom: the calibrated visibilities. }
\label{fig:redundant_baseline}
\end{figure}

In the Tianlai cylinder array, the receivers on the same cylinder are placed along the due North-South direction with equal spacings, 
though the spacing for the three cylinders are different (the three cylinders have 31, 32 and 33 
feeds respectively, each with a total length of 12.4 m, so that and the average 
center-to-center spacings are 0.4133 m, 0.40 m, 0.3875m respectively). Thus, for receivers on the 
same cylinder, except for the longest one, the baselines all have redundancy, with 
the shorter ones having more redundancy. 

In Fig.~\ref{fig:redundant_baseline} we show all the visibilities of the redundant baselines for a single frequency at the 
moment of Cyg A meridian transit.  If the gain of the receivers are the same, with only difference in the phase, we would expect
that the visibilities all have the same magnitude, but with different phases, so in the complex 
plane they should form a circle. As shown in the top panel of the figure, there is a circular distribution 
of the visibilities, but the magnitudes spread over the whole circle area, due to both differences in the receiver gain amplitude and the noise. 
The uncalibrated $V_0$ component extracted by the SPCA process is shown in the middle panel of Fig.~\ref{fig:redundant_baseline},
where the ring of data points have less spread in the radius, as the noise is removed.  Some points in the Origin are from malfunctioning 
feeds which produce too small output.
After calibration (bottom panel), for each redundant baseline,
the visibilities from the different pairs are indeed collapsed to a single point, and all the points of the different baselines form a nearly 
perfect circle, as one would have expected from the theory. This shows that our method of calibration indeed works with high precision. 
The ``redundant baseline" calibration method assums that the visibility from redundant baselines should all be the same, 
and uses this to solve for the array gain. However, noise or outliers may affect the precision of such calibration method, as shown by  
Fig.~\ref{fig:redundant_baseline}. If the redundant baseline calibration is performed when a strong source dominates, the 
SPCA method may also be applied to extract the signal component for use in the redundant baseline calibration, which may help improve the 
signal to noise ratio.

\subsection{Beam Profile}

\begin{figure}[htbp]
  \centering
  \subfloat[East-West pol\label{fig:ewp1}]{\includegraphics[width=0.42\textwidth]{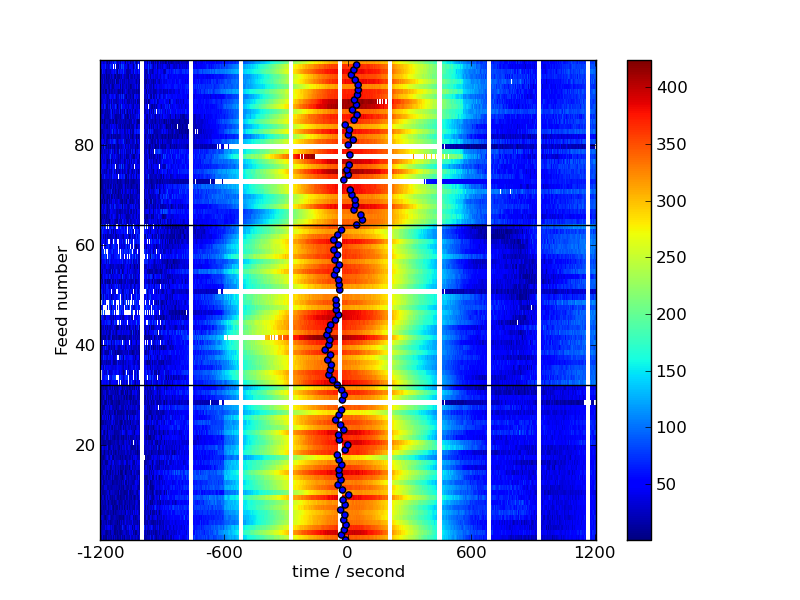}}\\
  \subfloat[North-South pol\label{fig:nsp1}]{\includegraphics[width=0.42\textwidth]{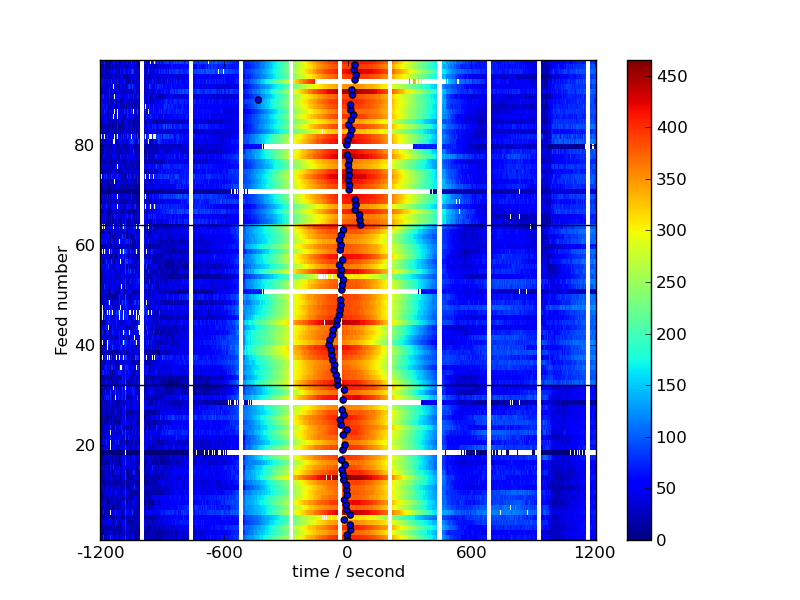}}
  \caption{The solved $|G|$ for the $XX$ and $YY$ polarization during 40 minutes of the Cyg A transit period. 
  The center of beam profiles is found by fitting a sinc function to the good data.}
  \label{fig:Gc}
\end{figure}

Fig.~\ref{fig:Gc} shows the  solved $|G_{i}|$ for all 96 feeds during the Cyg A transit, centering on the calculated time of astrometric
transit. We arrange the feeds on the three cylinders, which are clearly marked by the two dark horizontal lines in the figure. 
The regularly spaced vertical white stripes corresponds to the time of artificial calibrator broadcasting. 
The bad feeds are shown as blue/white horizontal lines.
From the figure, we can see also the approximate beam profiles for 
each feed, because $| G_{i} | \propto A_i(\hat{\boldsymbol{n}}_0)$ if
$g_{i}$ is constant or changes slowly.

During the construction of the array we made our best effort to install the feeds along the same North-South line and adjusted their 
pointing to be along the plumb line. However, it was understood that there maybe errors in both the manufacture and the installation of 
the feeds, and also winds etc. may affect position and pointing of the feeds. The cylinder reflector surface may also have some error.
As shown in Fig.~\ref{fig:Gc}, the measured beam profiles for the different feeds are not completely aligned; 
this is especially obvious in the second (middle) and third (top) cylinders. 
We flag out the abnormal ones, and then fit the remaining ones with a 
Gaussian function or a sinc function along the  east-west direction. The center point of the profile for the
sinc function fit is plotted in Fig.~\ref{fig:Gc} as the blue points in the center,
from which the mis-alignment of the beam is more apparently shown.  The maximum deviation of the transit peak is 108
seconds, corresponding to an angle of 0.45\degree. The median value of the deviation is 28 seconds, corresponding to an angle of 0.12\degree.
Field inspection and experiment is needed to determine the actual cause of the misalignment, which is beyond the scope of the 
present paper and will be investigated in future works on the testing of the Tianlai array.

\begin{figure}[tbp]
  \centering
  \subfloat[XX polarization, FWHM $= 3.6\degree$\label{fig:ewp3}]{\includegraphics[trim=30 10 30 30,clip,width=0.4\textwidth]{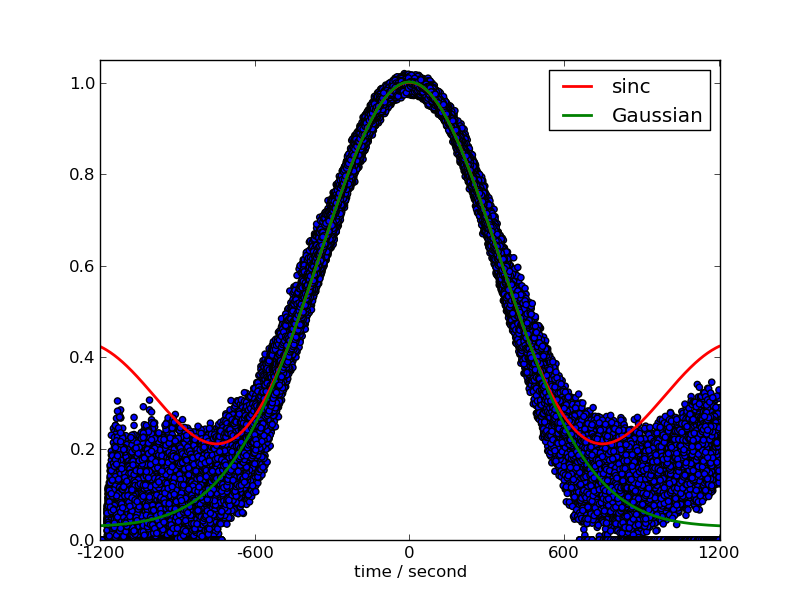}}\\
  \subfloat[YY polarization, FWHM $= 3.15\degree$\label{fig:nsp4}]{\includegraphics[trim=30 10 30 30,clip,width=0.4\textwidth]{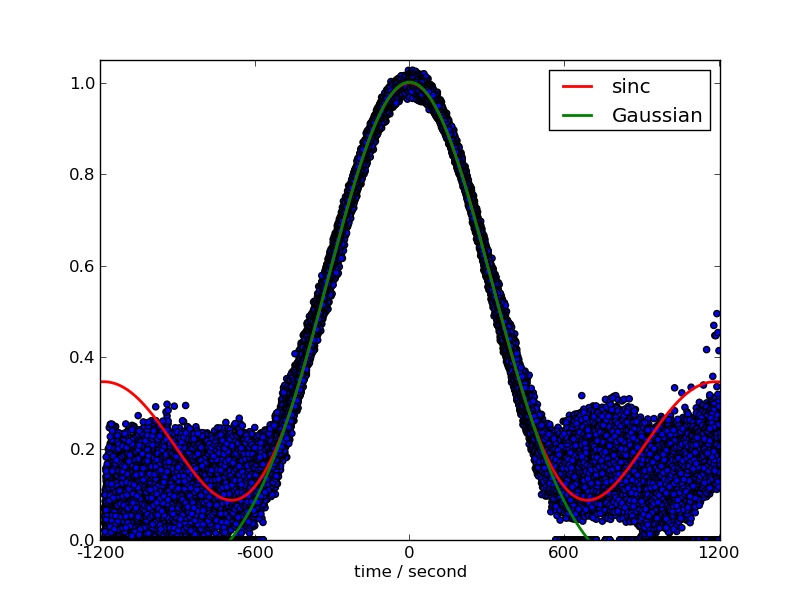}} 
  \caption{The fitted beam profile.}
  \label{fig:Bf}
\end{figure}

We fit a common beam profile by combining the normalized and
aligned data (exclude the bad/abnormal ones). The result for the frequency of $750 \MHz$ is shown in
Fig.~\ref{fig:Bf}, where we plot both the Gaussian and the sinc function fitting curves. 
The different receiver units have almost identical beam profile in the central part. In the 
side lobes the profiles of different units vary a lot, but note that in the side lobe there is also large
measurement error, as the calibrator signal is no longer dominant over the noise. The Gaussian and sinc
fitting curves also coincide with each other in the center part.  From 
the fitted Gaussian function we can obtain the FWHM of the beam width. Using 
$\theta_{\rm FWHM}=\Omega_{\rm Earth} \Delta T_{\rm FWHM}/ \cos\delta_{\rm src}$ ($\cos\delta_{\rm Cyg A}\approx 1$) 
we find the FWHM beam width is $3.6 \degree$ for the $XX$ polarization, and $3.15 \degree$ for the
$YY$ polarization.

\section{Conclusion}\label{S:con}

We have developed a method for the initial calibration of the complex gains of a 
radio interferometer array by taking the observational data of a strong point source, 
and arranging the visibilities (interferometer correlations) as a matrix $V_{ij}$ with indices denoting the pairs of receiver feeds, 
then solving for the eigenvector of the matrix. The eigenvector of the matrix with the largest eigenvalue gives a least
square solution to the complex gains of the receivers. To deal with the noise and outliers (e.g. malfunctioning feeds, and residual RFIs)
which are frequently seen in such data, we improve the method by first using a stable principal component analysis (SPCA) 
algorithm to decompose the visibility matrix into the point calibrator signal (a low rank matrix), an outlier component (a sparse matrix), and
a noise component (a matrix with dense small elements). When the calibrator signal is strong, this decomposition yields unique
solution. While in this paper we have applied the method to transit observations,  it can also be used for 
tracking observation. The method can also be extended to treat the calibration of full polarization responses, though in that
case calibration observation for at least three polarized calibrator sources are need for solving the additional parameters in the 
measurement equation. 

We applied this method to the first light data of the Tianlai cylinder pathfinder array. The calibration is performed using both 
periodically broadcasted artificial noise calibrator for the relative instrument phases and strong astronomical radio sources for 
both phase and amplitude of complex gains. We find that the 
instrument phases are very stable during the night,  though during day time the phases vary
as the environment temperature changes.  Checking with visibilities of the redundant baselines, we find 
that as expected, the calibrated visibilities form a circle on the complex plane, while the raw visibilities spread out as an irregular disk.   
The SPCA algorithm can be used to extract the signal component from the noise and outliers, which may also be useful to help 
improve the signal-to-noise ratio in the calibrations based on the redundant baselines. 
 
Based on the strong source transit data, the cylinder beam profile is measured along the East-West direction for each feed.
We find that despite engineering efforts,  there is some misalignment in the feed response, the exact cause is still to be determined. 
We also have fitted the beam profile with Gaussian and sinc functions. After adjusting for the misalignment, the central part of the beam for 
the different feeds agree very well, and the FWHM beam width are measured.

Much further analysis with more data is necessary to fully characterize the performance of the Tianlai array and to accurately 
calibrate its response. The aim of the present work is to present a method of array calibration based on 
eigenvector analysis and SPCA decomposition. The method is shown to work with a sample of the Tianlai data.  
We have incorporated this method in the Tianlai data processing pipeline\footnote{\url{https://github.com/TianlaiProject/tlpipe}}, and it will 
be used in our subsequent works on the testing and commissioning of the Tianlai pathfinder arrays.

\acknowledgments
The computations of this work was performed on the Tianhe-2 supercomputer at the National Supercomputing
Center in Guangzhou, Sun Yat-Sen University with the support of NSFC supercomputing Joint Grant U1501501.
The Tianlai survey is supported by the  MoST grant 2016YFE0100300.
This work is supported by the NSFC grant 11473044, 11633004, 
the NSFC-ISF joint research program No. 11761141012, 
and the CAS Frontier Science Key Project QYZDJ-SSW-SLH017. This document was 
prepared by the Tianlai collaboration using the resources
of the Fermi National Accelerator Laboratory (Fermilab), a U.S. Department
of Energy, Office of Science, HEP User Facility. Fermilab is managed by
Fermi Research Alliance, LLC (FRA), acting under Contract No.
DE-AC02-07CH11359.

\bibliographystyle{aasjournal}
\bibliography{cal}

\end{document}